\documentclass[aps,prd,showpacs,nofootinbib]{revtex4-1}

\usepackage{epsfig}
\usepackage{graphicx}
\usepackage{amsfonts}
\usepackage{amssymb}
\usepackage{bm}
\usepackage{latexsym}
\usepackage{amsmath}
\usepackage{amsbsy}
\usepackage{color}
\usepackage{comment}

\newcommand{\beq}{\begin{equation}}
\newcommand{\eeq}{\end{equation}}
\newcommand{\be}{\begin{equation}}
\newcommand{\ee}{\end{equation}}
\newcommand{\bea}{\begin{eqnarray}}
\newcommand{\eea}{\end{eqnarray}}

\newcommand{\1}{{\'{\i}}}

\def\1{\'{\i}}
\def\phidot{\dot\phi}

\def\half{\frac{1}{2}}

\def\thetadot{\dot\theta}
\def\psibar{{\bar{\psi}}}

\begin{document}


\title{Effective chemical potential in spontaneous baryogenesis}

\author{Arnab Dasgupta}
\email{arnabdasgupta@protonmail.ch}
\affiliation{School of Liberal Arts, Seoul National University of Science and Technology, 
232 Gongneung-ro, Nowon-gu, Seoul, 139-743, Korea}

\author{Rajeev Kumar Jain}
\email{rkjain@iisc.ac.in}
\affiliation{Department of Physics, Indian Institute of Science, \\ Bangalore 560012, India }

\author{Raghavan Rangarajan}
\email{raghavan@ahduni.edu.in}
\affiliation
{School of Arts and Sciences,  Ahmedabad University, \\ Navrangpura, Ahmedabad 380009, India}

\date{\today}


\begin{abstract}

Models of spontaneous baryogenesis have an interaction term $\partial_\mu\theta j^\mu_B$  in the Lagrangian, where $j^\mu_B$ is the baryonic current and $\theta$ can be a pseudo-Nambu-Goldstone boson. Since the time component of this term, $\thetadot j^0_B$, equals $\thetadot n_B$ for a spatially homogeneous current, it is usually argued that this term implies a splitting in the energy of baryons and antibaryons thereby providing an effective chemical potential for baryon number. In thermal equilibrium, one {then obtains} $n_B \sim \thetadot T^2$.  We however argue that a term of this form in the Lagrangian does not contribute to the single particle energies of baryons and antibaryons.  We show this for both fermionic and scalar baryons.  But, similar to some recent work, we find that despite the above result the baryon number density obtained from a Boltzmann equation analysis  can be proportional to $\thetadot T^2$. Our arguments are very different from that in the standard literature on spontaneous baryogenesis.

\end{abstract}
\pacs{98.80.Cq}


\maketitle
\section{Introduction}

Observations indicate that our Universe possesses a baryon asymmetry. The conventional approach to baryogenesis in cosmology is based on the three well known (and necessary) Sakharov's conditions \cite{Sakharov:1967dj}: (i) violation of baryon number  (ii) violation of C- and CP-symmetries and (iii) being out of thermal equilibrium. However, there exist some interesting scenarios wherein one or more of these conditions are not satisfied. The spontaneous baryogenesis scenario is one such novel scenario in which the CP-symmetry is not violated and the baryon asymmetry is generated in thermal equilibrium. 

Models of spontaneous baryogenesis \cite{Cohen:1987vi,Cohen:1988kt} have an interaction of the form $\partial_\mu\theta j^\mu$ in the Lagrangian density, where $j^\mu$ is related to the baryonic current and $\theta$ may be a 
pseudo-Nambu-Goldstone boson. Now, $\int d^3x j^0=Q$, where $Q$ is the charge associated with $j^\mu$, and ignoring spatial variations in $j^0$, $j^0=Q/V= n$, where $n$ is the net number density of the quanta associated with scalars or fermions
$\phi$ or $\psi$.  The coefficient of $n$, i.e. $\thetadot$, has been interpreted to be equivalent to an energy splitting in particle and antiparticle energies and thus an effective chemical potential for $\phi$ or $\psi$, 
provided the rate of change of
$\dot\theta$ is sufficiently slow.  This can then give rise to a particle-antiparticle asymmetry in thermal equilibrium. This interpretation has been invoked in spontaneous baryogenesis, including at the electroweak phase transition, and in flat direction baryogenesis, radion baryogenesis, quintessential baryogenesis, etc.

In this article we question the arguments underlying the above interpretation.  We argue that a $\thetadot n$ term in the Lagrangian density does not necessarily imply a split in the energies of particles and antiparticles and hence does not automatically lead to an interpretation of $\thetadot$ being an effective chemical potential.  We also argue that dispersion relations $k^0({\bf k})$ do not necessarily give particle and antiparticle energies. For the latter one must obtain the Hamiltonian and take its expectation value in single particle and antiparticle states.  The energies one obtains do not always agree with the expressions for $k^0$.  In particular, while $k^0$ may contain $\thetadot$ the single particle and antiparticle energies may not.

For the models under discussion we include a baryon number violating interaction and  further study the Boltzmann equation, similar to the approach of Ref. \cite{Arbuzova:2016qfh}. For scenarios with a $\thetadot j^0$ term in the Lagrangian density, the dispersion relations are modified, but, interestingly, even for cases where single particle and antiparticle energies are the same one does get a net baryon asymmetry due to the modified dispersion relations. Depending on the baryon number violating term, one gets different  expressions for the asymmetry.  This mechanism of generation of asymmetry from a $\thetadot j^0$ term is very different from that originally proposed in spontaneous baryogenesis and similar scenarios.

Spontaneous baryogenesis models \cite{Cohen:1987vi,Cohen:1988kt} also consider the generation of baryon asymmetry in the oscillating phase of the
$\theta$ field.  This has been further commented upon in Refs. \cite{Dolgov:1994zq,Dolgov:1996qq} and we do not
consider this here.

The outline of our article is as follows. In Sections \ref{sec:fermions} and \ref{mynote} we discuss the case of a fermion current coupled to the derivative of a field $\theta$.  We obtain the dispersion relation and the single particle and antiparticle energies.  We then perform an analysis using the Boltzmann equation. In Sec. \ref{sec:scalars1} we consider the case of a scalar field with an interaction similar to that in Sec. II, i.e., a coupling of the scalar field current with $\partial_\mu\theta$. In Sec. \ref{sec:scalars2} we consider the case of a scalar field with a self interaction of the form $g_2 n$, where $g_2$ is a constant.  In both cases, as in the fermionic case, we obtain the dispersion relations and single particle and antiparticle energies and then perform an analysis using the Boltzmann equation. Finally, we summarize our conclusions in Sec. \ref{sec.conclusion}.


\section{Fermions and $\partial_\mu\theta j_{\psi}^\mu$}
\label{sec:fermions}

Let us first consider fermions $\psi$ coupled to a field $\theta$ as
\be
{\cal L}= i\psibar\gamma^\mu\partial_\mu \psi -m\psibar\psi + \half v^2 \partial_\mu\theta\partial^\mu \theta 
-{\partial_\mu \theta j_{\psi}^\mu} -V(\theta,\psi)
\label{lagrangian}
\ee
Here $\theta$ may be a pseudo-Nambu-Goldstone boson associated with the spontaneous breaking of some symmetry at a scale $v$, and {the fermionic current is} $j^\mu_\psi=\psibar\gamma^\mu\psi$. In the literature, it has been argued that the time component of {the interaction term} $\partial_\mu \theta j^\mu_\psi$ in the Lagrangian density is $\thetadot (n_\psi - n_\psibar)$,
if spatial variations in $j^0$ can be ignored, and so $\thetadot$ acts like an effective chemical potential for $\psi$, in that it gives contributions with different signs to the single particle energies of particles and antiparticles which would enter in a Fermi-Dirac distribution.  This would lead to a net asymmetry in $\psi$ if the interactions of 
$\psi$ that change $\psi$ number are in thermal equilibrium. Let us investigate this proposition.

In standard free fermion field theory one writes down a Lagrangian density for a free field $\psi$.  One can then expand
\begin{equation}
\psi({\bf x},t)=\int d^3k 
\left[
b_s({\bf k}) u_s({\bf k}) f(t) \exp(+i{\bf k.x}) + d_s^\dagger({\bf k}) v_s({\bf k}) g(t) \exp(-i{\bf k.x})\right] \,.
\label{psi}
\end{equation}
One  then substitutes this in the Euler-Lagrange equation, i.e., the Dirac equation, and obtains $f(t)= \exp(-i k^0t)$ and $g(t)=\exp(+ik^0t)$, where $k^0=\sqrt{{\bf k}^2 + m^2}$, after associating positive and negative `energy' solutions with spinors $u$ and $v$ respectively. One subsequently obtains solutions for  $u$ and $v$.
One then writes the Hamiltonian density ${\cal H}=p_\psi \dot\psi -{\cal L}$ and substituting the above expression for $\psi$ 
in $\cal H$, one finds that the eigenvalue, and expectation value, of the Hamiltonian for a one particle state is $k^0$.
It is at this stage that one makes the identification that $k^0=\sqrt{{\bf k}^2 + m^2}$ is the energy $E$
of the one particle state.

We follow the same logic for our study.  We start with a general expansion for the fermion field as in Eq. (\ref{psi}). Then we obtain the functions $f(t)$ and $g(t)$ from the Euler-Lagrange equation for the Lagrangian density in Eq. (\ref{lagrangian}). Spontaneous baryogenesis scenarios must include baryon $(\psi)$ number violating interactions (such as the  last term of Eq. (2.5) of Ref. \cite{Cohen:1988kt}) to generate a difference in particle-antiparticle number densities from a difference in particle-antiparticle energies.  Such interactions are not relevant for the discussion below and we ignore $V(\theta,\psi)$ in the equation of motion for the fermionic field/spinors.
The equation for $u$ is
\begin{equation}
[i\gamma^0\dot f/f-\gamma^i k^i-m-\thetadot\gamma^0]u=0\,,
\label{eq:udirac}
\end{equation}
where we have ignored spatial variations in $\theta$. Multiplying from the left by  
$[i\gamma^0\dot f/f-\gamma^i k^i-\thetadot\gamma^0+m]$, we get
\begin{equation}
(i\dot f/f-\thetadot)^2=k_i^2+m^2\equiv E_*^2\,.
\end{equation}
Keeping the positive square root on the r.h.s. above, i.e. the positive `energy' solution, $\dot f/f=-i(E_*+\thetadot)$ and so
\begin{equation}
f=e^{-i\int (E_*+\thetadot) dt}\equiv e^{-i\int k^0_u dt} \,,
\label{f}
\end{equation}
where $k^0_u=E_*+\thetadot\,.$ Similarly we get
\be
[i\gamma^0\dot g/g+\gamma^i k^i-m-\thetadot\gamma^0]v=0\,,
\label{eq:vdirac}
\ee
and  multiplying from the left by 
$[i\gamma^0\dot g/g+\gamma^i k^i-\thetadot\gamma^0+m]$ we get
\begin{equation}
(i\dot g/g-\thetadot)^2=k_i^2+m^2\equiv E_*^2.
\end{equation}
Now, keeping the negative `energy' solution, $\dot g/g= i(E_*-\thetadot)$ and so we get
\begin{equation}
g=e^{+i\int (E_*-\thetadot) dt}\equiv e^{+i\int k^0_v dt} \,,
\label{g}
\end{equation}
with $k^0_v=E_*-\thetadot\,$. Now let us solve for $u$ and $v$.  If one puts the above expression for $f$ in Eq. (\ref{eq:udirac}) then the $\thetadot$ cancels out and the equation for $u$ is
\be
[E_*\gamma^0-\gamma^i k^i -m]u=0 \,.
\label{ueom2}
\ee
Similarly for $v$ we get
\be
[E_*\gamma^0-\gamma^i k^i + m]v=0 \,.
\label{veom2}
\ee
These are the standard equations for the spinors with solutions in the Dirac-Pauli representation as
\begin{align}
u_s(k)
& =  \alpha \begin{pmatrix} \widetilde{u}_s\\ \frac{{\bf \sigma}.{{\bf k}}}{(E_*+ m )}\widetilde{u}_s\end{pmatrix} \nonumber \\
v_s(k) 
& = \beta \begin{pmatrix} \frac{{\bf \sigma}.{\bf{k}}}{(E_* +m)}\widetilde{v}_s \\ \widetilde{v}_s\end{pmatrix} \,,
\end{align}
where 
$\widetilde{u}_1=\widetilde{v}_2=
\begin{pmatrix}
1\\0
\end{pmatrix}$ and
$\widetilde{u}_2=\widetilde{v}_1=
\begin{pmatrix}
0\\1
\end{pmatrix}$.
$\widetilde{u}^\dagger_{s'}\widetilde{u}_s=\widetilde{v}^\dagger_{s'}\widetilde{v}_s = \delta_{ss'}$. $\sigma_i$ are the Pauli matrices.  We will first determine the normalisation constants $\alpha$ and $\beta$ using commutation relations.

Considering the field expansion in Eq. (\ref{psi}), we can express $b_s({\bf k})$ and $d_s({\bf k})$ in terms of  the field $\psi({\bf x},t)$ as follows.
\begin{align}
\int d^3 x \,e^{-i{\bf k}'.{\bf x}} \psi({\bf x},t) &= \sum_{s}\int d^3x\int d^3k\left[b_s({\bf k})u_s({\bf k})f(t)e^{i({\bf k}-{\bf k}').{\bf x}} + d^{\dagger}_s(-{\bf k}) v_s(-{\bf k})g(t)e^{i({\bf k}-{\bf k}').{\bf x}}\right] \nonumber \\
&= \sum_{s} \int d^3 k\, (2\pi)^3 \delta({\bf k}-{\bf k}') \left[b_s({\bf k})u_s({\bf k})f(t)+d^{\dagger}_s(-{\bf k})v_s(-{\bf k})g(t)\right]\,.
\label{eq:psi}
\end{align}
Multiplying   Eq. \eqref{eq:psi} by $u^{\dagger}_{s'}({\bf k'})$ we get
\begin{align}
\int d^3 x \,e^{-i{\bf k}'.{\bf x}} u^\dagger_{s'}({\bf k}')\psi({\bf x},t) &= \sum_{s} \int d^3 k\, (2\pi)^3 \delta({\bf k}-{\bf k}') \left[b_s({\bf k})u^\dagger_{s'}({\bf k}') u_s({\bf k})f(t)+d^{\dagger}_s(-{\bf k})u^\dagger_{s'}({\bf k}')v_s(-{\bf k})g(t)\right]\nonumber \\
&= b_{s'}({\bf k'}) (2\pi)^3 |\alpha|^2 f(t)\frac{2E_*}{E_*+m}.
\end{align}
Therefore
\begin{align}
b_s({\bf k}) &= \frac{E_* + m}{2(2\pi)^3E_* |\alpha|^2f(t)}\int d^3 x \,e^{-i{\bf k}.{\bf x}} u^{\dagger}_s ({\bf k}) \psi({\bf x},t)\,.
\label{eq:bs}
\end{align}
Similarly, we obtain that
\begin{align}
d^{\dagger}_s({\bf k}) &= \frac{E_* + m}{2(2\pi)^3E_* |\beta|^2g(t)}\int d^3 x\, e^{i{\bf k}.{\bf x}} v^\dagger_s ({\bf k})  \psi({\bf x},t) \,.
\label{eq:ds}
\end{align}
From Eqs. \eqref{eq:bs} and \eqref{eq:ds} we also get
\begin{align}
b^{\dagger}_s({\bf k}) &= \frac{E_* + m}{2(2\pi)^3E_* |\alpha|^2f^*(t)}\int d^3 x \,e^{i{\bf k}.{\bf x}} \psi^{\dagger} ({\bf x},t)  u_s({\bf k})
\label{eq:bds}
\end{align}
and
\begin{align}
d_s({\bf k}) &= \frac{E_* + m}{2(2\pi)^3E_* |\beta|^2g^*(t)}\int d^3 x \,e^{-i{\bf k}.{\bf x}} \psi^\dagger ({\bf x},t)  v_s({\bf k}) \,.
\label{eq:dds}
\end{align}
We further note the equal time commutation relation
\begin{align}
\bigg{\{}\psi({\bf x},t),\Pi_\psi({\bf y},t)\bigg{\}} &= i\delta({\bf x-y})\,,
\label{eq:commpsiPi}
\end{align}
where
$\Pi_\psi=i\psibar\gamma^0$, implies
\begin{align}
\bigg{\{}\psi({\bf x},t),\psi^{\dagger}({\bf y},t)\bigg{\}} &= \delta({\bf x-y})\,.\label{eq:compsi}
\end{align}
Then using  Eqs. \eqref{eq:bs}, \eqref{eq:bds} and \eqref{eq:compsi}, we get
\begin{align}
\{b_s({\bf k}),b^{\dagger}_{s'}({\bf k}')\} &= \frac{(E_* + m)^2}{4(2\pi)^6|\alpha|^4E^{2}_*} \int d^3 x \int d^3 y\, e^{-i{\bf k}.{\bf x}+i{\bf k}'.{\bf y}} u^\dagger_s({\bf k})  \{\psi({\bf x},t) ,\psi^\dagger({\bf y},t)\}  u_{s'}({\bf k}') \nonumber \\
&=\frac{(E_* + m)}{2(2\pi)^6|\alpha|^2 E_{*}} (2\pi)^3 \delta({\bf k}-{\bf k}') \delta_{ss'},
\label{eq:bbd}
\end{align}
where in the last step we have used ${\bf k}={\bf k}'$. Similarly,
\begin{align}
\{d_s({\bf k}),d^{\dagger}_{s'}({\bf k}')\} &= \frac{(E_* + m)}{2(2\pi)^6 |\beta|^2 E_* } (2\pi)^3 \delta({\bf k}-{\bf k}')\delta_{ss'}.
\label{eq:ddf}
\end{align}
Now,  demanding the commutation relations
\begin{align}
\{b_s({\bf k}),b^{\dagger}_{s'}({\bf k}')\} &= (2\pi)^3 \delta({\bf k}-{\bf k}')\delta_{ss'} \nonumber \\
\{d_s({\bf k}),d^{\dagger}_{s'}({\bf k}')\} &= (2\pi)^3 \delta({\bf k}-{\bf k}')\delta_{ss'}
\end{align}
we get
\be
|\alpha|^2 = \frac{E_* + m }{2(2\pi)^6 E_*},
\hspace{1cm}
|\beta|^2 = \frac{E_* + m}{2(2\pi)^6 E_*}\,.
\ee
One can also show that $\{b_s({\bf k}),d_{s'}({\bf k}')\}$ and $\{b^\dagger_s({\bf k}),d^\dagger_{s'}({\bf k}')\}$ are 0.
Further imposing 
\begin{equation}
\bigg{\{}\psi({\bf x},t),\psi({\bf y},t)\bigg{\}} = 
\bigg{\{}\psi^\dagger({\bf x},t),\psi^\dagger({\bf y},t)\bigg{\}}=
0
\label{eq:commpsipsi}
\end{equation}
allows us to show that all other commutation relations involving the annihilation and creation operators, such as
$\{b_s({\bf k}),d^\dagger_{s'}({\bf k}')\}$, etc., are 0.

We now obtain the Hamiltonian density from the Lagrangian density as ${\cal H}=\sum_\varphi p_\varphi \dot\varphi - \cal L$,  where $\varphi$ represents the fermionic and the $\theta$ fields.  
$\Pi_\psi=i\bar\psi\gamma^0$.
 Assuming $\theta$ has no other time derivative couplings $\Pi_\theta= v^2\thetadot- \psibar\gamma^0\psi$.  Then the Hamiltonian density is
\bea
{\cal H} &=& \Pi_\psi \dot\psi + \Pi_\theta\thetadot - {\cal L}\nonumber\\
&=&-i\psibar\gamma^i\partial_i\psi + m\psibar\psi 
+\half v^2 \thetadot^2 +\half v^2 (\nabla \theta)^2
+\partial_i \theta\psibar\gamma^i\psi
+ V(\theta,\psi)
\label{Hftheta}
\eea
One might now conclude, from the form of the Hamiltonian density in Eq. (\ref{Hftheta}), that particles and antiparticles have the same energy and that the $\thetadot$ term does not lead to energy splitting.
\footnote{A similar argument was made in the arXiv version of Ref.  \cite{Dolgov:1996qq}.}
But then one could argue that one should write the Hamiltonian in terms of $\Pi_{\theta}$ and not $\dot\theta$ and that gives a $(\Pi_\theta/v^2) j^0$ term in the Hamiltonian.
This might suggest that energies of particles and antiparticles may also depend on $\Pi_\theta$. But since one is 
arguing in terms of an (effective) {\it chemical potential} giving rise to a net asymmetry in thermal equilibrium
one should consider situations involving the 
Fermi-Dirac distribution $[1+\exp(-(H-\mu)/T)]^{-1}$ or
Boltzmann factor, where $\mu$ is the 
chemical potential associated with conserved charges, while
effective chemical potential terms, such as those possibly associated with $\dot\theta$,
come, if at all, from $H$ as per the arguments of spontaneous baryogenesis.
Now we would like to draw an analogy with a particle of charge $q$ moving with velocity $\bf v$ in a magnetic field 
$\bf{B}$.  
The Hamiltonian for the charged particle written in terms of the canonical momentum $\bf p$ is
\be
H= \frac{({\bf p} - q {\bf A}/c)^2}{2m}\,,
\ee
which seems to indicate a difference of $q\, {\bf p.A}/(mc)$ in the energies of particles and antiparticles.
However 
when one uses the Boltzmann factor $\exp[-(H-\mu)/T]$ and integrates it over the momentum components $(p_x,p_y,p_z)$ from 
$-\infty$ to $+\infty$ to obtain
the number of particles and antiparticles, 
one can change the integration variables from 
$(p_x, p_y, p_z)$ to the kinetic momentum components
$(k_x, k_y, k_z)=(p_x-qA_x/c, p_y-qA_y/c, p_z-qA_z/c)$, also varying from $-\infty$ to $+\infty$.
Then though $q \bf{A}/c$ is part of the Hamiltonian when written in terms of the canonical momentum, 
it does not play a role in determining the number of particles in statistical physics.  Thus
even if the Hamiltonian written in terms of the canonical momentum seems to indicate a difference in particle and antiparticle energies 
it does not necessarily lead to an energy splitting and a net asymmetry in thermal equilibrium. 
For our case, one may argue that the Hamiltonian can be written initially in terms of $\Pi_\theta$ and one can
perform a change of variable, $\dot\theta=(\Pi_\theta+\psibar\gamma^0\psi)/v^2$, in which case the 
$(\Pi_\theta/v^2) j^0$ term will disappear and the Hamiltonian density will contain $\half v^2 \thetadot^2$ and not lead to a splitting
in the energy of fermions and antifermions.  
(Just as the energy density of the magnetic field does not
contribute to fixing the number density of a gas of charged particles and antiparticles in a magnetic field, 
$\half v^2 \thetadot^2$, the kinetic energy density of the $\theta$ field in our case, will
not contribute to setting the number density of fermions and antifermions.)

To resolve the matter of the energy of particles and antiparticles in the presence of the $\theta$ field we shall
obtain the expectation value of the Hamiltonian for single particle and antiparticle states.
The fermionic Hamiltonian (ignoring spatial variation in $\theta$) is (see Appendix \ref{sec.appendix1} for details)
\begin{eqnarray}
H &= &
\sum_{s,s'}
\int d^3 {k} \, (2\pi)^3
\left[
\overline{u}_s({\bf k})(\gamma^i  k_i + m) u_{s'}({\bf k}) b^{\dagger}_s({\bf k})b_{s'}({\bf k}) + \overline{v}_s({\bf k})(-\gamma^{i}k_i + m)v_{s'}({\bf k}) d_s({\bf k})d^{\dagger}_{s'}({\bf k}) \right]
\nonumber \\
&= &
\sum_{s}
\int d^3 {k} \, (2\pi)^3 \frac{2 E^{2}_*}{(E_* + m)}\bigg( |\alpha|^2 
b^{\dagger}_s({\bf k})b_{s}({\bf k}) - |\beta|^2 d_s({\bf k}) d_s^{\dagger}({\bf k})\bigg).
\label{eq:ham_ferm}
\end{eqnarray}
Now, using Eq. \eqref{eq:ddf}, the normal ordered fermionic Hamiltonian  becomes 
\begin{equation}
:H: \ =\  \sum_{s}\int \frac{d^3 {k}}{(2\pi)^3} \left[b^{\dagger}_s({\bf k})b_{s}({\bf k}) + d^{\dagger}_{s}({\bf k})d_{s}({\bf k})\right]\sqrt{{\bf k}^2 + m^2}\,.
\label{eq:finalfermionhamiltonian}
\end{equation}
The only $\thetadot$ dependence in the fermionic field is in $f$ and $g$ but, as in the standard case, the only terms above that survive go as $f^*f$ and $g^*g$ and so the $\thetadot$ dependence drops out.  Therefore the eigenvalue, and
expectation value, of the Hamiltonian is 
$\sqrt{{\bf k}^2 + m^2}$ for a fermion or an antifermion state and does not contain $\dot\theta$.  
The above calculation also underscores the point that in this case the $k_{u,v}^0$ in the exponent of $f,g$, which do contain  
$\dot\theta$, are not to be identified with fermionic energies,  that is, even if the dispersion relations $k^0({\bf k})$ contain $\thetadot$ the fermionic energies do not.  

We realize that it is inconsistent to keep the $\thetadot$ term and ignore other interactions of $\psi$ in the equation of motion for the fermionic field/spinors.  However, here we are merely trying to point out that the identification of $\thetadot$ with a difference in
fermion-antifermion energies is not justified. 

One may argue that in some scenarios one treats the $\theta$ field as an external field and so one may not define $\Pi_\theta$ or include $\Pi_\theta\thetadot$ in the Hamiltonian, and then there would be an energy splitting $\sim \thetadot$ between fermions and antifermions. We would however argue that dynamical versus external refers to whether the given Lagrangian/Hamiltonian 
determines the evolution equation of a field, i.e. its Euler-Lagrange equation, or whether there are other interactions that we are ignoring that determine it (this would perhaps be most relevant when the field is to be treated as a background classical field). It is in the context of the equation of motion for $\theta$ that one may differentiate between dynamical and external fields.  Now, going
from the Lagrangian to the Hamiltonian does not involve the equation of motion for $\theta$. So while obtaining the Hamiltonian from the Lagrangian we can and should follow the same approach irrespective of whether $\theta$ is a dynamical or an external field,
and we will then obtain the same result that there is no particle-antiparticle energy splitting.  
Another way of viewing this is that any external field is actually governed by a larger Lagrangian in which it is `dynamic' and so one should include the $\Pi_\theta\thetadot$ contribution while obtaining our Hamiltonian, since it is part of the larger Hamiltonian.  If the given Lagrangian or Hamiltonian does not include all interactions of $\theta$ then it will not be suitable for obtaining the equation of motion for $\theta$.\footnote{A discussion of dynamical and external fields is also included in Ref. \cite{Arbuzova:2016qfh}.} 


\subsection{Effective chemical potential in early Universe scenarios}

The discussion above is relevant for models of spontaneous baryogenesis. In the spontaneous baryogenesis mechanism \cite{Cohen:1987vi,Cohen:1988kt}, there is a term such as $\partial_\mu\theta j^\mu$ in the Lagrangian density, where $j^\mu$ is
the baryon current. (In some later models, $j^\mu$ is a current not orthogonal to baryon number, such as hypercharge.)  
In Refs. \cite{Cohen:1987vi,Cohen:1988kt} it was argued that treating $\theta$ as a classical background field the term $\thetadot j^0_B=\dot\theta n_B=\dot\theta (n_b-n_{\bar b})$ in the Lagrangian density,  where $n_{b,\bar b}$ are the baryon and antibaryon number densities, contributes differently to the energy of baryons and antibaryons.  Thus, $\dot\theta$ can be treated as an effective chemical potential for baryon number.  If this is the case then the net equilibrium baryon density in the thermal bath can be given by
the expression: $n_B^{eq}=g \dot\theta T^2/6$, where $g$ is the number of internal degrees of freedom of the baryons.   In the original spontaneous baryogenesis papers, which were not realised in the context of the electroweak phase transition, $\theta$ was a
slowly rolling pseudo-Nambu-Goldstone boson associated with the breaking of a $U(1)$ symmetry \cite{Cohen:1987vi,Cohen:1988kt}. When these models were applied to the electroweak phase transition, the role of $\theta$ was played by either a singlet field or a Higgs doublet \cite{Dine:1990fj,Dine:1991ck}, or by a phase associated with the Higgs doublets or hypercharge rotation in a 2-Higgs doublet model \cite{Cohen:1991iu,Bonini:1996xa}.\footnote{In models of spontaneous baryogenesis at the electroweak phase
transition, sphaleron processes are not fast enough to give thermal distributions for baryons and antibaryons.  Therefore, one uses
$\dot n_B=-\Gamma_B(n_B-n_B^{eq})\approx \Gamma_B n_B^{eq}$ for $n_B\ll n_B^{eq}$.  $\Gamma_B$ is the rate of B violation and $n_B^{eq}$ is taken to be $\sim \mu_B T^2$, with $\mu_B$, the effective chemical potential for baryon number, set equal to $\dot\theta$.}
In Ref. \cite{Cohen:1994ss} the effects of diffusion were included. In Ref. \cite{Abel:1992za} spontaneous baryogenesis was considered in the Minimal Supersymmetric Standard Model (MSSM) and an extension, and in Ref. \cite{Comelli:1993ne} spontaneous baryogenesis was considered in the MSSM with spontaneous CP violation. In Ref. \cite{Reina:1993ws} spontaneous baryogenesis in a left-right symmetric model was studied. In Ref. \cite{Giudice:1993bb} the effect of strong sphalerons on reducing the baryon asymmetry generated at the electroweak phase transition was discussed.
Our comments above on the identification of $\dot\theta$ as an effective chemical potential for baryon number or some related charge are valid for these scenarios.

There have been other attempts at calculating the baryon asymmetry generated at the electroweak phase transition with a 
$\thetadot j^0$ term in the Lagrangian density but without identifying $\thetadot$ with a chemical potential, such as  in Refs. \cite{Comelli:1994di, Joyce:1994fu, Joyce:1994zt, Bhatt:2004cq, Bhatt:2004zk}.
Our comments do not apply to these works. (In the classical force mechanism of electroweak baryogenesis of Refs. \cite{Joyce:1994fu,Joyce:1994zt,Bhatt:2004cq, Bhatt:2004zk} there is a $\partial_\mu \theta j^\mu_5$ term in the Lagrangian density.  But by going into the frame of reference of the expanding Higgs bubble wall one can eliminate the $\partial_0\theta$ term.  Then our arguments above are not relevant. In Ref. \cite{Cohen:1994ss} the final calculations are done in the bubble wall frame and our comments do not apply there. For the discussion in the previous sub-section the $\thetadot$ term in the Lagrangian density can not be removed by a change of reference frame.)

There have been several adaptations of the original idea of spontaneous baryogenesis using a term like $\thetadot j^0$ term in the Lagrangian density to generate the matter-antimatter asymmetry of the Universe.  Our arguments above on the identification of 
$\dot\theta$ with an effective chemical potential are also applicable to these scenarios. 
In Refs. \cite{Chiba:2003vp,Takahashi:2003db} the role of $\theta$ is played  by a slowly moving field associated with a SUSY flat direction while in Ref. \cite{Alberghi:2003ws}, the radion field in a braneworld setup assumes the role of a dynamical field. 
Models of quintessential baryogenesis \cite{Li:2001st, DeFelice:2002ir, Li:2002wd, Yamaguchi:2002vw} create a matter-antimatter asymmetry by coupling the slowly moving quintessence field to the baryon or lepton current, in a manner similar to models of spontaneous baryogenesis. 
Furthermore, in Refs. \cite{Davoudiasl:2004gf,Li:2004hh} a gravitational interaction between the 
time derivative of the Ricci scalar (or a function of it) and the baryon-number current has been studied. Such a coupling together with  baryon-number-violating interactions is used to produce an observationally acceptable baryon asymmetry in equilibrium.  
In Ref. \cite{Carroll:2005dj} the baryon asymmetry is generated using the time derivative of a scalar field coupled to the baryon current,  where the scalar field is a ghost field, a quintessence field and a pseudo-Nambu Goldstone boson.
In certain models of asymmetric dark matter \cite{Banks:2006xr, MarchRussell:2011fi,Kamada:2012ht}, the asymmetry in the dark matter which is linked to that in baryons, is generated by an effective chemical potential.
In Refs. \cite{Kusenko:2014lra,Pearce:2015nga,Yang:2015ida,Kusenko:2014uta} the time variation of the Higgs field or the axion field as it relaxes in the early Universe to 0 from a large value obtained during inflation is treated as an effective chemical potential for $B+L$ or some  fermionic number. 
In Ref. \cite{Daido:2015gqa} the time dependent axion field at a point as an axion domain wall passes through it is treated as an effective chemical potential, based on the derivative coupling of the axion field to the Standard Model left-handed lepton current.
In Ref. \cite{Takahashi:2015ula} a current-current interaction in the Lagrangian density implies that the zeroth component of the current associated with a complex inflaton field $\phi$, $i(\dot{\phi} \phi^* - \phi\dot{\phi^*})$, is coupled to a $B-L$ charge density, and it is presumed to act as a chemical potential for $B-L$.
Ref. \cite{DeSimone:2016juo} considered spontaneous baryogenesis driven by a non-canonical scalar field.

One may consider a coupling of fermions with another field that does not involve derivative couplings, such as  the QED Lagrangian density with the term $-q\psibar\gamma^\mu\psi A_\mu= -q A_0 (n_\psi-n_\psibar) -q \psibar\gamma^i\psi A_i$.  We are interested in the first term.  Since there is no time derivative it will appear in the Hamiltonian density with a positive sign, and imply that charged particles and antiparticles have opposite sign energies in an electric field. 
If $A_\mu$ represents  a background field one may treat $A_0$ as an effective chemical potential for the fermions.
If $A_\mu$ is not a physical field but simply a spacetime dependent function, or a constant vector, $A_0$ may again be treated as an effective chemical potential.

\section{Boltzmann equation}
\label{mynote}

In Ref. \cite{Arbuzova:2016qfh} the authors have also discussed the identification of $\thetadot$ with an effective chemical potential.  They study the kinetic equation for baryons and find that while the naive interpretation of $\thetadot$ as an effective chemical potential is not appropriate, surprisingly the kinetic equation indicates that the baryon number density is dependent on $\thetadot$ in
a way that $c\thetadot$ plays a role similar to that of an effective chemical potential, where $c$ is a constant whose value may be 
different for different types of B-nonconserving reactions.   Our analysis below is similar to that in Ref.  \cite{Arbuzova:2016qfh} but the fermions in our Lagrangian density
are transformed so that the Lagrangian density contains a term   $(\partial_\mu \theta) J^\mu$, where $J^\mu$ is the baryonic current, as in models of spontaneous
baryogenesis.

In Ref. \cite{Arbuzova:2016qfh} one considers the Lagrangian density of a complex scalar field $\Phi$ interacting with fermions $Q_1$ and $L$.
\be
{\cal L }(\Phi,Q,L) =  g^{\mu\nu} \partial_\mu \Phi^*
\partial_\nu \Phi - V(\Phi^* \Phi) + \bar Q_1 (i \gamma^\mu \partial_\mu - m_Q)\,Q_1 
+  \bar L ( i \gamma^\mu \partial_\mu - m_L) L + {\cal L}_{int}(\Phi, Q_1, L)\, ,
\label{S-Phi}
\ee
where  $Q_1$ and $\Phi$ have nonzero baryonic numbers 1/3 and -1, while $L$ does not carry baryonic charge. 
\be
{\cal L}_{int} =  \frac{\sqrt 2}{m_X^2} \frac{\Phi}{v}\, (\bar L \gamma_{\mu} Q_1 )(\overline {Q_1^c}  \gamma_{\mu} Q_1) +
h.c. \, , 
\label{L-int}
\ee
where $Q_1^c$ is a charge conjugated quark spinor and $m_X$ and $f$ are parameters with dimensions of mass.
\be
V(\Phi^* \Phi) = \lambda \left(\Phi^* \Phi - v^2/2 \right)^2  .
\label{V-of-Phi}
\ee

After spontaneous symmetry breaking in the $\Phi$ sector, and ignoring the heavy radial mode and any phase in the vacuum expectation value of $\Phi$, $\Phi\rightarrow v e^{i\theta}/\sqrt2$, and the Lagrangian density is
\be
{\cal L}_1 (\theta,Q,L)
= \frac{v^2}{2} \partial_\mu \theta \partial^{\mu} \theta + \bar Q_1
(i \gamma^{\mu} \partial_\mu - m_Q) Q_1 +  \bar L (i \gamma^{\mu} \partial_\mu - m_L) L +
 \left(\frac{e^{i \theta }}{ m_X^2} \, (\bar L \gamma_{\mu} Q_1 )(\overline {Q_1^c}  \gamma_{\mu} Q_1) + h.c.\right) - U(\theta)\, ,
 \label{L-of-theta-1}
 \ee
where an explicit symmetry breaking term is included as $U(\theta)$.  Now, introducing a rotated field $Q_2$ through 
$Q_1 = e^{- i \theta /3} Q_2$, one gets
\bea
{\cal L}_2 (\theta,Q,L)
&=& \frac{v^2}{2} \partial_\mu \theta \partial^{\mu} \theta + \bar Q_2
(i \gamma^{\mu} \partial_\mu - m_Q) Q_2 +  \bar L (
i \gamma^{\mu} \partial_\mu - m_L) L  \nonumber \\
&&+ \left(\frac{1}{ m_X^2} \, 
(\bar L \gamma_{\mu} Q_2 )(\overline {Q_2^c}  \gamma_{\mu} Q_2)
+
  h.c.\right) + (\partial_\mu \theta) J^\mu - U(\theta) \, , 
  \label{L-of-theta-2}
 \eea
where the quark baryonic current is $J_\mu =(1/3) \bar Q_2 \gamma_\mu Q_2$.
 
We now consider the process $Q_2 Q_2 \leftrightarrow L \bar Q_2 $ $(1 2 \leftrightarrow 3 4)$.  Then
\begin{eqnarray}
\dot n_Q + 3 H n_Q &\sim& \int dt\,\int d\tau_{L \bar Q} \,d\tau_{QQ}  |A|^2 
\delta( {\bf k}_{in} -  {\bf k}_{out})
\exp[-it( k_1^0+k_2^0-k_3^0-k_4^0)] (f_3 f_4 - f_1 f_2)\label{bz1}\\
&=&\int  d\tau_{L \bar Q} \,d\tau_{QQ} 
|A|^2 \delta({\bf k}_{in} - {\bf k}_{out}) \delta( k_1^0+k_2^0-k_3^0-k_4^0) (f_3 f_4 - f_1 f_2)
\end{eqnarray}
where $A$ is the invariant amplitude, ${\bf k}_{in,out}$ refers to the total incoming and outgoing 3-momentum, and
$d\tau_{ab}=d^3k_a d^3k_b/[4E_a E_b (2\pi)^6]$ is the phase space factor.  
$k_{1,2}^0=E_{1,2}-\dot\theta/3$, $k_3^0=E_3$ and $k_4^0=E_4+\dot\theta/3$.  The $E_I$'s are physical energies and are $({\bf k}_I^2 + m_I^2)^\half$ while the $k_I^0$ appear in the expansion of the fermion field as in Eqs. (\ref{psi}), (\ref{f}) and (\ref{g}). 
(Because of a difference in the signs of the $\partial_\mu \theta J^\mu$ term, and of a factor of 1/3 in $J^\mu$, in Eq.
(\ref{L-of-theta-2}) and Eq. (\ref{lagrangian}) the expressions for $k_{1,2,4}^0$ and $k_{u,v}^0$ are a bit different.)
Note that to get the $k^0$ delta function $\thetadot$ has to be constant in time.\footnote{We would like to thank Prof. A. D. Dolgov for highlighting this to us.}
In the spontaneous baryogenesis scenario $\theta$ is slow moving compared to the time scale for particle interactions, and so one can take $\thetadot$ to be nearly constant. This requirement agrees with comments in Sec. IV-C of Ref. \cite{Arbuzova:2016qfh}.

The particle distribution functions are given by
\begin{equation}
f_I=\exp[-E_I/T+\xi_I]
\end{equation}
where $\xi_I=\mu_I/T$ and $\mu_I$ is the chemical potential of species $I$, and the antiparticle chemical potential is the negative of that of the particle (presuming fast annihilation of particle-antiparticle pairs into, say, photons).

Then one gets
\begin{eqnarray}
\dot n_Q + 3 H n_Q 
&\sim&\int  d\tau_{L \bar Q} \,d\tau_{QQ} 
|A|^2 \delta( {\bf k}_{in} -  {\bf k}_{out})\delta( E_1+E_2-E_3-E_4 -\dot\theta) (f_3 f_4 - f_1 f_2)\\
&=&\int  d\tau_{L \bar Q} \,d\tau_{QQ} 
|A|^2 \delta( {\bf k}_{in} -  {\bf k}_{out})\delta( E_1+E_2-E_3-E_4 -\dot\theta) \exp[-E_{in}/T]
\left(e^{\xi_3+\xi_4+\dot\theta/T}-e^{\xi_1+\xi_2}     \right),
\label{bz2}
\end{eqnarray}
where we have set $E_{in}\equiv E_1+E_2=E_3+E_4 +\dot\theta$ in the factor with the distribution functions, using the energy delta function. 
Following the analysis below  Eq. (4.15) of Ref. \cite{Arbuzova:2016qfh}, if $Q$ and $L$ are in thermal equilibrium the collision integral on the r.h.s of Eq. \eqref{bz2} vanishes and $\xi_3+\xi_4+\dot\theta/T=\xi_1+\xi_2$, or $\xi_L=3\xi_Q -\dot\theta/T$, assuming $Q-\bar Q$ annihilation processes to photons are in thermal equilibrium giving $\xi_{\bar Q}=-\xi_Q$.  
The interaction term in Eq. \eqref{L-int} is $B$ conserving but violates $B+L$, while that in Eqs.  (\ref{L-of-theta-1}) and (\ref{L-of-theta-2})
after symmetry breaking is $B$ violating but $B+L$ conserving.
Assuming that $B+L$ conserving processes, and not any $B+L$ violating processes, are in thermal equilibrium, and that initially $B+L$ is 0, $\xi_Q/3+\xi_L=0$. Combining these relations we get  $\xi_Q = 0.3\, \dot\theta/T$, and 
\bea
n_B&=&B_Q n_Q
=\frac{1}{6}g_Q B_Q \xi_Q T^3 
=\frac{1}{20} g_Q B_Q \thetadot T^2
= \frac{1}{30} \thetadot T^2\,
\label{eq:fermionbaryonnumber}
 \eea
where $g_Q$ is the number of  spin states of $Q$, namely 2, and $B_Q$ is the baryonic charge of $Q$.
\footnote{In Ref. \cite{Arbuzova:2016qfh} the expression  for the baryon number density $n_B$ in the rotated field case given above their Eq. (4.14) 
differs from that inferred from Eq. (4.17) for the unrotated case.  The expression for the rotated case has an error 
but by correctly considering the chemical potential for lepton number and imposing $B+L$ conservation the authors get the same result as for their unrotated case
 \cite{Arbuzova:2017private}, which also
agrees with our result
above. Our $n_Q$ and $\xi_Q$ are their $n_B$ and $\xi_B$.}

The above analysis shows that $\thetadot$  can play the role of a chemical potential for quarks, {\it but with a multiplicative factor} (such as 3/10 for the above case). But this does not follow simply from the argument that a $\thetadot j^0$ term in the Lagrangian density for fermions implies an energy splitting for fermions and antifermions.  In fact, as we have shown above, there is no split in the single particle energies of quarks and antiquarks in the presence of a non-zero $\thetadot$.
The multiplicative factor depends on what is the baryon number violating process in thermal equilibrium, and which charge, $B+L$ in our case, is conserved.  The former determines what are the {\it in} and {\it out} particles in the collision integral
and the form of the last bracket in  Eq. (\ref{bz2}), which is then set to 0. The latter gives another relation between chemical potentials.


\section{Scalars and $\partial_\mu\theta j^\mu$}
\label{sec:scalars1}

Let us now consider scalars coupled to another field as
\be
{\cal L}= \partial_\mu\phi^*\partial^\mu\phi - m^2\phi^*\phi + \half v^2 \partial_\mu\theta\partial^\mu \theta 
+\partial_\mu\theta j^\mu
-V(\theta,\phi)
\label{Lphi}
\ee
where $j^\mu=i(\phi^*\partial^\mu \phi-\partial^\mu\phi^*\,\phi)$ is the scalar current and its zeroth component is the scalar charge density, $n_\phi-n_{\phi^*}$, ignoring spatial variation in $j^0$.
The equation of motion from the above Lagrangian  density for $\phi^*$ and $\phi$ are given as
\begin{align}
\Box \phi^* + 2i \partial^\mu \theta \partial_\mu \phi^* + (m^2 + i \Box \theta) \phi^* &=0\label{eq:eomphic} \\
\Box \phi - 2i \partial^\mu \theta \partial_\mu \phi + (m^2 - i \Box \theta) \phi &=0\,.
\label{eq:eomphi}
\end{align}
We take the field expansion to be 
\begin{align}
\phi({\bf x},t) &= \int 
{d^3k}
\left[a({\bf k})f(t)e^{i{\bf k}.{\bf x}} 
+ b^{\dagger}({\bf k})g(t)e^{-i{\bf k}.{\bf x}}\right]\,.
\end{align}
Then the equation of motion for $\phi$, assuming $\theta$ is homogeneous, gives
\begin{align}
\ddot{f} + {\bf k}^2f - 2i\dot{\theta}\dot{f} + (m^2 -i\ddot\theta)f &= 0 
\end{align}
Taking $\ddot \theta \approx 0$, if $\theta$ is assumed to be slowly rolling, we get
\begin{align}
\ddot{f} - 2i\dot{\theta}\dot{f} + ( {\bf k}^2+ m^2)f &= 0 
\end{align}
Keeping the solutions that reduce to the standard positive and negative energy solutions in the absence of $\theta$, we get
\begin{align}
f(t) &= \alpha e^{-it( \sqrt{ {\bf k}^2 + m^2+\,\dot{\theta}^2} \,-\thetadot)} 
\equiv \alpha e^{-i k^0_1 t} 
\end{align}
and
\begin{align}
g(t) &= 
\beta e^{it( \sqrt{ {\bf k}^2 + m^2 +\, \dot{\theta}^2   }\,+\thetadot )}
\equiv \beta e^{i k^0_2 t}  \,,
\end{align}
where
\be
k_1^0=  \sqrt{ {\bf k}^2 + m^2 + \dot{\theta}^2 } -\thetadot = \sqrt{ E_*^2 + \dot{\theta}^2 } -\thetadot
\ee
and
\be
k_2^0=  \sqrt{ {\bf k}^2 + m^2 + \dot{\theta}^2 } +\thetadot = \sqrt{ E_*^2 + \dot{\theta}^2 } +\thetadot\,.
\ee
$E_*=({\bf k}^2 + m^2)^{\frac12}$, as before. Then
\begin{align}
\phi({\bf x},t) &= \int {d^3k}
\left[\alpha \,a({\bf k})  e^{i{\bf k}.{\bf x}-i k^0_1 t} 
+ \beta \,b^{\dagger}({\bf k})e^{-i{\bf k}.{\bf x} +i k^0_2 t}\right]\,.
\label{phiexp}
\end{align}
We impose the equal time commutation relations 
\begin{align}
\left[\phi({\bf x},t),\phi({\bf y},t)\right] &
= \left[\phi^*({\bf x},t),\phi^*({\bf y},t)\right] 
= \left[\phi({\bf x},t),\phi^*({\bf y},t)\right] = 0
\label{comm1} \\
\left[\Pi_\phi({\bf x},t),\Pi_\phi({\bf y},t)\right] &
= \left[\Pi_{\phi^*}({\bf x},t),\Pi_{\phi^*}({\bf y},t)\right] 
= \left[\Pi_\phi({\bf x},t),\Pi_{\phi^*}({\bf y},t)\right] = 0
\label{comm2} \\
\left[\phi({\bf x},t),\Pi_\phi({\bf y},t)\right] &
= \left[\phi^*({\bf x},t),\Pi_{\phi^*}({\bf y},t)\right] 
=i\delta({\bf x}-{\bf y})
\label{comm3}
\end{align}
where
\be
\Pi_\phi=\frac{\partial {\cal L}}{\partial \dot\phi}=\phidot^*+i\thetadot\phi^*\qquad{\rm and}\qquad
\Pi_{\phi^*}=
\frac{\partial {\cal L}}{\partial \dot\phi^*}=
\phidot-i\thetadot\phi\,.
\ee
Eq. \eqref{comm3} and the last equality of Eq. \eqref{comm2} imply
\begin{align}
\left[\phi({\bf x},t),\dot{\phi}^*({\bf y},t)\right] &= 
\left[\phi^*({\bf x},t),\dot{\phi}({\bf y},t)\right] = i\delta({\bf x} - {\bf y}) \\
\left[\dot{\phi}^*({\bf x},t),\dot{\phi}({\bf y},t)\right] &= 2\dot{\theta} \,\delta({\bf x} - {\bf y}).  
\end{align}

Taking linear combinations of $\phi$ and $\dot\phi$, and of their complex conjugates, the annihilation and creation operators can be written as
\begin{align}
a({\bf k}) &= \frac{i}{\alpha (k_1^0+k_2^0)}\int \frac{d^3x}{(2\pi)^{3}}\left(\dot{\phi}({\bf x},t) - i k_2^0\phi({\bf x},t)\right)e^{-i({\bf k}.{\bf x} - k_1^0 t)} \\
b({\bf k}) &= \frac{i}{\beta^* (k_1^0+k_2^0)}\int \frac{d^3x}{(2\pi)^{3}}\left(\dot{\phi}^*({\bf x},t) - i k_1^0\phi^*({\bf x},t)\right)e^{-i({\bf k}.{\bf x} - k_2^0 t)} \\
a^\dagger({\bf k}) &= \frac{-i}{\alpha^* (k_1^0+k_2^0)}\int \frac{d^3x}{(2\pi)^{3}}\left(\dot{\phi}^*({\bf x},t) + i k_2^0\phi^*({\bf x},t)\right)e^{i({\bf k}.{\bf x} - k_1^0 t)} \\
b^\dagger({\bf k}) &= \frac{-i}{\beta (k_1^0+k_2^0)}\int \frac{d^3x}{(2\pi)^{3}}\left(\dot{\phi}({\bf x},t) + i k_1^0\phi({\bf x},t)\right)e^{i({\bf k}.{\bf x} - k_2^0 t)} \,.
\end{align}
Then imposing the commutation relations
\begin{align}
[a({\bf k}),a^{\dagger}({\bf k}')] &= (2\pi)^3 \delta({\bf k}-{\bf k}') \nonumber \\
[b({\bf k}),b^{\dagger}({\bf k}')] &= (2\pi)^3 \delta({\bf k}-{\bf k}')
\end{align}
we get
\be
|\alpha|^2 = |\beta|^2=\frac{1}{(2\pi)^6}\frac{1 }{2(E_*^2+\thetadot^2)^\frac{1}{2}} \,.
\ee
One can also show that $[a({\bf k}),b({\bf k}')]=[a^\dagger({\bf k}),b^{\dagger}({\bf k}')]=0$.
Expressing $\phi$ as $\phi=1/{\sqrt 2}[\phi_1+i\phi_2]$ and imposing
\begin{equation}
\left[\phi_i({\bf x},t),\Pi_{i}({\bf y},t)\right] 
=i\delta({\bf x}-{\bf y})
\end{equation} implies
\begin{equation}
\left[\phi({\bf x},t),\dot{\phi}({\bf y},t)\right] = 
\left[\phi^*({\bf x},t),\dot{\phi}^*({\bf y},t)\right]=0\,.
\end{equation}
Then the first 2 commutation relations in Eq. (\ref{comm2}) imply
\begin{equation}
\left[\dot\phi({\bf x},t),\dot{\phi}({\bf y},t)\right] = 
\left[\dot\phi^*({\bf x},t),\dot{\phi}^*({\bf y},t)\right]=0\,.
\end{equation}
One can then show that all other commutation relations involving the annihilation and creation operators, such as
$[a({\bf k}),b^{\dagger}({\bf k}')]$, etc., are 0.

Assuming $\theta$ has no other derivative couplings, $\Pi_\theta= v^2\thetadot + i(\phi^* \phidot-\phidot^*\,\phi)$ and the Hamiltonian density is
\bea
{\cal H} &=& \Pi_\phi \dot\phi + \Pi_{\phi^*}\dot\phi^*+
\Pi_\theta\thetadot - {\cal L}\\
&=&\phidot^*\phidot + |\nabla\phi|^2 + m^2\phi^*\phi
+\half v^2 \thetadot^2 
+i\thetadot (\phi^*\phidot-\phidot^*\,\phi)
+ V(\theta,\psi)
\label{Hthetaphi}
\eea
Unlike in the fermionic case, here $\thetadot$ appears in the Hamiltonian with $j_0$ and so looks like an effective chemical potential.  
Now one may believe that in this case one will get a $\thetadot$ splitting in the energy.  But  the normal ordered Hamiltonian of the 
terms involving $\phi$ gives
\bea
:H:&=&\,: \int \frac{d^3k}{(2\pi)^3}
\left[
\left((k_1^{0})^2 + {\bf k}^2 +m^2 + 2\thetadot k_1^0\right)a^\dagger({\bf k}) a({\bf k})
+\left((k_2^{0})^2 + {\bf k}^2 +m^2 - 2\thetadot k_2^0\right)b({\bf k}) b^\dagger({\bf k})
\right.\nonumber
\\
& & \qquad\qquad \left.
\quad -2\thetadot^2(a^\dagger({\bf k})b^\dagger(-{\bf k})+a({\bf k})b(-{\bf k})
\right]
\frac{1}{2(E_*^2+\thetadot^2)^\frac{1}{2}}
:\nonumber\\
& = &
\int \frac{d^3k}{(2\pi)^3}\,
[a^\dagger({\bf k}) a({\bf k}) + b^\dagger({\bf k}) b({\bf k})] 
\frac{E_*^2}{(E_*^2+\thetadot^2)^\frac{1}{2}}
-\left[(a^\dagger({\bf k})b^\dagger(-{\bf k})+a({\bf k})b(-{\bf k})\right]
\frac{\thetadot^2}{(E_*^2+\thetadot^2)^\frac{1}{2}}
\,.
\eea
The particle and antiparticle energies $E$ are defined as the expectation value of the Hamiltonian in single particle and antiparticle states (rather than eigenvalues of the Hamiltonian because of the terms in the second bracket above).  They are the same and are given by
 \be
 E=\frac{E_*^2}{(E_*^2+\thetadot^2)^\frac{1}{2}}
 =\frac{{\bf k}^2+m^2}{({\bf k}^2 + m^2+\thetadot^2)^\frac{1}{2}} \,.
 \ee
It is not at all obvious that the Hamiltonian in Eq. (\ref{Hthetaphi}) will give no energy splitting between particles and antiparticles.  Naively, one would have concluded the opposite.

We now consider an interaction $g\phi^*\phi\chi^*\phi + h.c.$ and do a Boltzmann equation analysis as in Sec. \ref{mynote} for the process $\phi^*\, \phi\leftrightarrow \chi\phi^* (12\leftrightarrow34)$.
We assume $\phi$ has baryon number equal to 1 and $\chi$ has lepton number equal to 1. 
Starting from 
\be
\dot n_B + 3 H n_B \sim \int dt\,\int d\tau_{\chi \phi^*} \,d\tau_{\phi^*\phi} |A|^2 
\delta( {\bf k}_{in} -  {\bf k}_{out})
\exp[-it( k_1^0+k_2^0-k_3^0-k_4^0)] (f_3 f_4 - f_1 f_2)\,,
\label{bz-scalar1}
\ee
we find that the time integral gives
\bea
\delta \left(k_1^0+k_2^0-k_3^0-k_4^0\right)
&=&\delta\left(\sqrt{E_{*1}^2+\thetadot^2}+\sqrt{E_{*2}^2+\thetadot^2}
-E_3-\sqrt{E_{*4}^2+\thetadot^2 } -\thetadot\,\right)\nonumber\\
&=&\delta\left(E_1\frac{E_{*1}^2+\thetadot^2}{E_{*1}^2}
+E_2\frac{E_{*2}^2+\thetadot^2}{E_{*2}^2}
-E_3
-E_4\frac{E_{*4}^2+\thetadot^2}{E_{*4}^2} -\thetadot
\right)\,.
\eea
 This does not give a neat relation one can use to substitute in the
 distribution functions in Eq. \eqref{bz-scalar1}.
 
 If we assume $\thetadot$ is small compared to $E_{*I}$ then we get
 \begin{eqnarray}
 \!\!\!\!
\dot n_B + 3 H n_B 
&\sim&\int  d\tau_{\chi \phi^*} \,d\tau_{\phi^*\phi} 
|A|^2 \delta( {\bf k}_{in} -  {\bf k}_{out})\delta( E_1+E_2-E_3-E_4 -\dot\theta) (f_3 f_4 - f_1 f_2)\nonumber\\
&=&\int  d\tau_{\chi \phi^*} \,d\tau_{\phi^*\phi} 
|A|^2 \delta( {\bf k}_{in} -  {\bf k}_{out})\delta( E_1+E_2-E_3-E_4 -\dot\theta) \exp[-E_{in}/T]
\left(e^{\xi_3+\xi_4+\dot\theta/T}-e^{\xi_1+\xi_2}     \right),
\end{eqnarray} 
In thermal equilibrium the collision integral vanishes and, assuming $\phi-\phi^*$ annihilation processes are in thermal equilibrium, 
$\xi_{\phi^*}=-\xi_\phi$.
Then $\xi_\chi=\xi_\phi-\thetadot/T$.  Further assuming that $B+L$ conserving processes are in thermal equilibrium, 
and that initial $B+L$ was zero, $\xi_\chi=-\xi_\phi$.  Then $\xi_\phi=(1/2)\thetadot/T$ and
\be
n_B=B_\phi (n_\phi-n_{\phi^*})=\frac{1}{3}\xi_\phi T^3=
\frac{1}{6}\thetadot T^2\,.
\ee
Again, $\thetadot$ with a multiplicative factor has played the role of a chemical potential for $\phi$.  More precisely, for this particular case, $\thetadot/2$ is the chemical potential for $\phi$.  But there is no split in the single particle energies of $\phi$ particles and antiparticles, though unlike in the fermionic case they are amended, equally, by the presence of $\thetadot$.

\section{Scalars with $g_2 j^0$}
\label{sec:scalars2}

In Ref. \cite{Carcamo:2012cv} the authors considered a Lagrangian that is the field theoretic equivalent of a charged particle in a magnetic field, in a scenario motivated by the coupling of a string to an external magnetic field.  The relevant term in the Lagrangian density reduces to a form $g_2 j^0$, where $g_2$ is a constant.  The analysis in Ref. \cite{Carcamo:2012cv} was in 1+1 dimensions and used the difference in the energies of particles and antiparticles to obtain a matter-antimatter symmetry.

Here we consider a scalar field with a similar Lagrangian density as in Ref. \cite{Carcamo:2012cv} but in $3+1$ dimensions.
The Lagrangian density is given by
\begin{equation}
{\cal L} = \frac{1}{2} \left((\dot{\phi}_{i})^{2} - (\phi'_{i})^{2}\right) - \frac{g_{2}}{2} \epsilon_{ij} \dot{\phi}_{i} \phi_{j}.\label{L2}
\end{equation}
We are interested in studying the effects on the energies of particles and antiparticles due to the presence of such an interaction term. The canonical conjugate momenta corresponding to the fields $\phi_{i}$ are given by
\begin{equation}
\Pi_{i} (t,{\bf x}) = \frac{\partial {\cal L}}{\partial \dot{\phi}_{i}} = \dot{\phi}_{i}  -\frac{g_{2}}{2} \epsilon_{ij} \phi_{j}.
\label{M2}
\end{equation}
Using this, we obtain the Hamiltonian density as 
\begin{equation}
{\cal H} = \Pi_{i} \dot{\phi}_{i} - {\cal L} = \frac{1}{2}\left((\dot{\phi}_{i})^{2} + (\phi'_i)^{2}\right).\label{H2}
\end{equation}
Rewriting the Lagrangian density in Eq. (\ref{L2}) in terms of $\phi^{\pm} = \frac{1}{\sqrt{2}}(\phi_{1} \pm i\phi_{2})$, one gets
\begin{equation}
{\cal L}= \phidot^+ \phidot^- - \phi^{+\prime} \phi^{-\prime} -\frac{g_2}{2i}(\phidot^-\phi^+-\phidot^+\phi^-).
\end{equation}
Using  $\phi^+=\phi^{-*}\equiv\phi$, the Lagrangian density becomes
\begin{equation}
{\cal L}= \phidot \phidot^{*} -\phi^{\prime} \phi^{\prime*} 
-\frac{g_2}{2}i(\phi^{*}\phidot -\phi \phidot^{*}).
\label{Lphig}
\end{equation}
The last term is  $-(g_2/2) j^0$ and is equivalent to $-(g_2/2) (n_\phi-n_{\phi^*})$, ignoring spatial variation in $j^0$.  The Lagrangian density is similar to that in Eq. \eqref{Lphi} but with no mass term or dynamical $\theta$ field and with $\partial_\mu\theta j^\mu$ or $\thetadot j^0$ replaced by $-(g_2/2) j^0$.
Then the field $\phi$ can be expanded as in Eq. \eqref{phiexp} with
\be
k_{1,2}^0=\sqrt{{\bf k}^2+\frac{g_2^2}{4}}\pm \frac{g_2}{2}
\label{disprelationsk0g2}
\ee
and
\be
|\alpha|^2 = |\beta|^2=\frac{1}{(2\pi)^6}\frac{1 }{2({\bf k}^2+{g_2^2}/{4})^\frac{1}{2}}
\,.
\ee

Now, Eq. (\ref{Lphig}) gives
\be
{\cal H} =  \Pi_\phi \phidot + \Pi_{\phi*}\phidot^* -{\cal L} = \left(\dot\phi^* \phidot + \phi^{*'}\phi'\right),\label{Hphi}
\ee
where
\be
\Pi_\phi=\phidot^*+\frac{g_2}{2i}\phi^* \quad {\rm and} \quad \Pi_{\phi*}=\phidot- \frac{g_2}{2i}\phi\,,
\ee
and $\prime$ refers to the spatial derivative.

One might now conclude, from the form of the Hamiltonian density in Eq. (\ref{Hphi}), that particles and antiparticles have the same energy and that the $g_2$ term does not lead to energy splitting.  
But, as argued in Sec. \ref{sec:fermions}, one should not derive such conclusions from the form of the Hamiltonian density. 

One might also argue, as in Ref. \cite{Carcamo:2012cv}, that the dispersion relations in Eq. \eqref{disprelationsk0g2} imply an energy splitting for scalars and antiscalars. But, again, as seen earlier in Secs \ref{sec:fermions} and \ref{sec:scalars1}, $k^0$ and particle energies may differ, and a difference in $k^0$ for particles and antiparticles does not necessarily lead to a difference
in particle and antiparticle energies.

So, as before, we now explicitly calculate single particle/antiparticle energies.
The normal ordered Hamiltonian can be written (ignoring the $a^\dagger b^\dagger$ and $ab$ terms) as
\begin{equation}
:H:\,= \int  
\frac{d^3k}{(2\pi)^3}
\left(\frac{(k_1^0)^2+{\bf k}^2}{2({\bf k}^2+{g_2^2}/{4})^\frac{1}{2}} a_{\bf k}^\dagger a_{\bf k} + 
\frac{(k_2^0)^2+{\bf k}^2}{2({\bf k}^2+{g_2^2}/{4})^\frac{1}{2}} b_{\bf k}^\dagger b_{\bf k}\right).
\end{equation}
Taking expectation values in single particle and antiparticle states one can immediately infer that the energy of a particle is 
\begin{eqnarray}
E=\frac{(k_1^{0})^2 + {\bf k}^2} {2({\bf k}^2+{g_2^2}/{4})^\frac{1}{2}}
\approx 
|{\bf k}|+\frac{g_2}{2}\,,
\label{Ep}
\end{eqnarray}
if $g_2\ll |{\bf k}|$, while the energy of an antiparticle is
\begin{eqnarray}
E=\frac{(k_2^{0})^2 + {\bf k}^2} {2({\bf k}^2+{g_2^2}/{4})^\frac{1}{2}}
\approx 
|{\bf k}|-\frac{g_2}{2}\,.
\label{Eap}
\end{eqnarray}
So the single  particle energies of the particle and antiparticle are $g_2$ dependent and different.

We again consider an interaction $g\phi^*\phi\chi^*\phi + h.c.$, with $\phi$ and $\chi$ carrying baryon and lepton number of 1
respectively, and do a Boltzmann equation analysis as in Sec. \ref{sec:scalars1} for the process 
$\phi^*\, \phi\leftrightarrow \chi\phi^* (12\leftrightarrow34)$. The time integral in Eq. \eqref{bz-scalar1} will give
\bea
\delta \left(k_1^0+k_2^0-k_3^0-k_4^0\right)
&=&\delta\left(\sqrt{{\bf k}_{1}^2+g_2^2/4}+\sqrt{{\bf k}_{2}^2+g_2^2/4}
-E_3-\sqrt{{\bf k}_{4}^2+g_2^2/4 } + g_2/2\,\right)\nonumber\\
&\approx&\delta( E_1+E_2-E_3-E_4 + g_2/2)
\,.
\eea
Then we get
 \begin{eqnarray}
 \!\!\!\!\!\!\!\!
\dot n_B + 3 H n_B 
&\sim&\int  d\tau_{\chi \phi^*} \,d\tau_{\phi^*\phi} 
|A|^2 \delta( {\bf k}_{in} -  {\bf k}_{out})\delta( E_1+E_2-E_3-E_4 +g_2/2) (f_3 f_4 - f_1 f_2) \nonumber \\
&=&\int  d\tau_{\chi \phi^*} \,d\tau_{\phi^*\phi} 
|A|^2 \delta( {\bf k}_{in} -  {\bf k}_{out})\delta( E_1+E_2-E_3-E_4 +g_2/2) e^{-E_{in}/T}
\left(e^{\xi_3+\xi_4-g_2/(2T)}-e^{\xi_1+\xi_2}  \right).
\end{eqnarray} 
In thermal equilibrium the collision integral vanishes and, assuming $\phi-\phi^*$ annihilation processes and $B+L$ conserving processes are in thermal equilibrium, and that initial $B+L$ was zero, $\xi_\phi=-g_2/(4T)$. 
Both $\xi_\phi$ and the energy splitting of $g_2$ between particle and antiparticle energies will contribute to $(n_\phi- n_{\phi^*})$ to give
\be
n_B=B_\phi (n_\phi- n_{\phi^*})=
-\frac{g_2}{4}\thetadot T^2\,,
\ee
where the effective chemical potential for $\phi$ is $-g_2/4-g_2/2= -3g_2/4$.  
Unlike in the other cases studied in this article,  the effective chemical potential gets a contribution from the split in the single particle
energies of $\phi$ particles and antiparticles.

Ref. \cite{Carcamo:2012cv} obtains an asymmetry by only considering the difference in energies of particles and antiparticles with $\xi_\phi$ set to 0. Now $\xi_\phi$ depends on the $B$ violating process that is in equilibrium and creates the asymmetry between particles and antiparticles.
If ${\cal L} \supset g \phi^2 \chi^*\chi + h.c.$ then processes like $\phi \phi \rightarrow \chi \chi^*$ or $\phi \chi \rightarrow \phi^* \chi$ in equilibrium will give $\xi_\phi=0$, and the asymmetry in $\phi$ and $\phi^*$ will only be because of the energy splitting.

\section{Discussion and Conclusion}
\label{sec.conclusion}

Spontaneous baryogenesis presumes that a term of the form $\thetadot j^0 \sim \thetadot n$ in the Lagrangian density, where $j^0$ is the zeroth component of the particle current and $n$ is the net particle number density, translates into a splitting of  energies of particles and antiparticles, and therefore acts as an effective chemical potential which then gives rise to a  
matter-antimatter asymmetry or baryon asymmetry in a system in thermal equilibrium.
Our analysis above implies that there are two separate issues here.  One is whether or not the term in the Lagrangian density gives
rise to an energy splitting, and the second is whether or not one obtains a matter-antimatter asymmetry.  For both fermions and scalars we find in Secs. \ref{sec:fermions} and \ref{sec:scalars1} that $\thetadot j^0$ modifies the mode functions of the corresponding quantum fields, but it does not lead to a splitting of single particle and antiparticle energies.  However, because of the modified mode functions, if baryon number violating interactions are in the thermal equilibrium then equating the collision integral on the r.h.s. of the Boltzmann equation to 0 gives a non-zero chemical potential for particle number.  This then gives rise to a matter-antimatter asymmetry, or baryon asymmetry if the particles carry baryon number.

It may be noted that obtaining the Hamiltonian density in terms of the field and its time derivative and trying to relate the presence or absence of a $\thetadot j^0$ in the Hamiltonian density with particle-antiparticle energy splitting is inappropriate and misleading.  
For the fermionic case there is no $\thetadot j^0$ term in the Hamiltonian density while it does appear for the scalar case.  
But in both cases the single particle energies for particles and antiparticles are the same. 

The presence or absence of $\thetadot j^0$ in the Hamiltonian density does not depend whether or not $\theta$ is a dynamical field or an external field (like a classical background field).  In the latter case, the equation of motion of the field is not entirely determined by the Lagrangian under study but one still includes $\Pi_\theta \thetadot$ in the Hamiltonian density.  Therefore our conclusions above are the same irrespective of whether $\theta$ is a dynamical or an external field.

The presumption that a term of the form $\thetadot j^0 \sim \thetadot n$ in the Lagrangian density gives rise to a splitting in energies of particles and antiparticles has been widely used in the literature in models of spontaneous baryogenesis, including at the electroweak phase transition, flat direction baryogenesis, radion baryogenesis, quintessential baryogenesis, etc.
Our analysis above indicates that this presumption may not hold even though such a term may ultimately give rise to a matter-antimatter asymmetry.

We have also studied the case where the field derivative $\thetadot$ is replaced by a constant, in Sec. \ref{sec:scalars2}.  This modifies the mode functions and gives a splitting in energies of particles and antiparticles.  Then both the energy splitting and any chemical potential, depending on the number violating interactions in thermal equilibrium, will contribute to the asymmetry in number densities of particles and antiparticles. These conclusions would also hold for a term of the form $h(\theta) j^0$.

We conclude by comparing our results with that of Ref. \cite{Arbuzova:2016qfh} for the case of a fermionic current coupled with the derivative of the $\theta$ field.
In  Eq. (4.3) of Ref. \cite{Arbuzova:2016qfh} the energies of particles and antiparticles are obtained  from the dispersion relation.  We argue that the dispersion relation gives
expressions for $k^0$ which enter in the mode functions of the fermion field $\psi$ (see our Eqs. (\ref{f}) and (\ref{g})) but the single 
particle and antiparticle energies are given by the expectation value of the Hamiltonian in Eq. (\ref{eq:finalfermionhamiltonian}) in single 
particle and antiparticle states.  We too find that the $k^0$ associated with $u$ and $v$ spinors depend on $\dot\theta$ and are different, 
but our Hamiltonian implies that particles and antiparticles have the same energy which is independent of $\dot\theta$.  Our 
analysis  of the collision integral in the Boltzmann equation in Sec. \ref{mynote}
was done for the case where the quark field is rotated
to absorb a phase factor of $\exp(i\theta)$ after spontaneous symmetry breaking which gives rise to the $(\partial_\mu \theta) J^\mu$
in the Lagrangian density.  In Ref. \cite{Arbuzova:2016qfh} it was done for the unrotated case and our result for the baryon asymmetry
generated in thermal equilibrium for the rotated quark fields agrees with their result.  Ref. \cite{Arbuzova:2016qfh} also obtains a result for the baryon
asymmetry for the rotated case using other arguments. 
This has an error but after correction it is in agreement with that for the unrotated case and our result (see the footnote below our Eq.  (\ref{eq:fermionbaryonnumber})).

\begin{acknowledgments}
The authors would like to thank Elena Arbuzova, Ashok Das, A. D. Dolgov, Katherine Freese, Arul Lakshminarayan, Hiranmaya Mishra and Mark Srednicki
for useful comments and discussions.  AD, RKJ and RR would like to thank the organisers of the Workshop in High Energy Physics and Phenomenology (WHEPP13) at Puri, India where this work was reinitiated.
\end{acknowledgments}

\appendix 
\section{The fermionic Hamiltonian}
\label{sec.appendix1}
The fermionic Hamiltonian (ignoring spatial variation in $\theta$) is
\begin{eqnarray}
H &\equiv& \int d^3 x\, {\cal H} \nonumber \\
& = & \sum_{s,s'} \int d^3 {k} \, (2\pi)^3 \left[
\overline{u}_s({\bf k})(\gamma^i  k_i + m) u_{s'}({\bf k}) b^{\dagger}_s({\bf k})b_{s'}({\bf k}) + \overline{v}_s({\bf k})(-\gamma^{i}k_i + m)v_{s'}({\bf k}) d_s({\bf k})d^{\dagger}_{s'}({\bf k}) \right] \nonumber \\
& = &  \sum_{s,s'} \int d^3 {k}\,  (2\pi)^3 \left[
|\alpha|^2 \bigg{(} \tilde{u}_s^{\dagger}, \; -\tilde{u}^{\dagger}_s \frac{{\bf \sigma} \cdot {\bf k}}{E_* +m}\bigg{)} \begin{pmatrix} \left(m+\frac{{\bf k}^2}{E_* +m}\right) \tilde{u}_{s'} \\ 
\left(-1 + \frac{m}{E_* + m} \right)\sigma \cdot {\bf k} \,\tilde{u}_{s'}
\end{pmatrix} b^{\dagger}_{s'}({\bf k}) b_s({\bf k}) \right. \nonumber \\ 
& & \qquad \qquad \quad \qquad +  \left. |\beta|^2 \bigg{(} \tilde{v}^{\dagger}_s \frac{\sigma \cdot {\bf k}}{E_* +m},\; -\tilde{v}^{\dagger}_s \; \bigg{)} \begin{pmatrix} -\left(1 + \frac{m}{E_* + m} \right){\bf \sigma} \cdot {\bf k} \,\tilde{v}_{s'} \\ \left(m+\frac{{\bf k}^2}{E_* +m}\right) \tilde{v}_{s'}
\end{pmatrix} d_s({\bf k})d^{\dagger}_{s'}({\bf k})\right] \nonumber \\
&= & \sum_{s,s'}
\int d^3 {k} \,
(2\pi)^3
\left[
|\alpha|^2\, \frac{2m^3 + 2m^2E_* + 2{\bf k}^2(E_* + m)}{(E_* + m)^2}\,\delta_{ss'}b^{\dagger}_s({\bf k})b_s({\bf k})  
\right. \nonumber \\
& & \qquad \qquad \quad \qquad + \left. |\beta|^2 \,\frac{-2m^3 - 2m^2E_* - 2{\bf k}^2(E_* + m)}{(E_* + m)^2}\, \delta_{s s'}d_s({\bf k})d^{\dagger}_s({\bf k})  \right]\nonumber \\
&= &\sum_{s}
\int d^3 {k} \, (2\pi)^3\left(
2 |\alpha|^2 \frac{(m^2 + {\bf k}^2)(E_* + m)}{(E_* + m)^2} b^{\dagger}_s({\bf k})b_{s}({\bf k}) - 2 |\beta|^2 \frac{(m^2 + {\bf k}^2)(E_* + m)}{(E_* + m)^2} d_s({\bf k})d^{\dagger}_{s}({\bf k})\right) \nonumber \\
&= &
\sum_{s}
\int d^3 {k} \, (2\pi)^3 \frac{2 E^{2}_*}{(E_* + m)}\bigg(
|\alpha|^2 
b^{\dagger}_s({\bf k})b_{s}({\bf k}) - |\beta|^2 d_s({\bf k}) d_s^{\dagger}({\bf k})\bigg).
\end{eqnarray}
Upon using Eq. \eqref{eq:ddf}, we then obtain 
\begin{equation}
H = \sum_{s}\int d^3 {k} \,
(2\pi)^3 \frac{2E^{2}_*}{(E_* + m)}
\bigg(
|\alpha|^2  b_s^{\dagger}({\bf k}) b_s({\bf k}) + |\beta|^2  d_s^{\dagger}({\bf k})d_{s}({\bf k})\bigg)
- \sum_{s}\int d^3 {k} E_*\delta(0_{\bf k}).
\end{equation}
With this, our normal ordered fermionic Hamiltonian  becomes 
\begin{align}
:H: &= \sum_{s} \int 
\frac{d^3 {k}}{(2\pi)^3}  
\left(
\frac{E_* + m}{2E_*}\frac{2E^{2}_*}{(E_* + m)} b^{\dagger}_{s}({\bf k})b_{s}({\bf k}) + \frac{E_* + m}{2E_*}\frac{2E^{2}_*}{(E_* + m)} d^{\dagger}_{s}({\bf k})d_{s}({\bf k}) 
\right)
\nonumber \\
&= \sum_{s}\int \frac{d^3 {k}}{(2\pi)^3}   \left[b^{\dagger}_{s}({\bf k})b_{s}({\bf k}) +  d^{\dagger}_{s}({\bf k})d_{s}({\bf k})\right]E_*\nonumber \\
&= \sum_{s}\int \frac{d^3 {k}}{(2\pi)^3} \left[b^{\dagger}_s({\bf k})b_{s}({\bf k}) + d^{\dagger}_{s}({\bf k})d_{s}({\bf k})\right]\sqrt{{\bf k}^2 + m^2}\,.
\end{align}

\bibliographystyle{JHEP}
\bibliography{mf-bgenesis-refs}

\end{document}